\documentclass[11pt,a4paper,fleqn]{article}

\usepackage{amsmath,amssymb,amsthm,enumerate,cite}

\newtheorem{theorem}{Theorem}

\newtheorem{lemma}{Lemma}
{\theoremstyle{definition}
\newtheorem{definition}{Definition}
\newtheorem{note}{Note}
\newtheorem*{note*}{Note}
}

\begin{document}
\begin{center}
{\LARGE\bf Conservation Laws and Potential Systems of Diffusion-Convection Equations\\[1.3ex]}
{\large Nataliya~M.~Ivanova\\[1.7ex]}
\footnotesize
Institute of Mathematics of National Academy of Sciences of Ukraine, \\
3, Tereshchenkivska Str., Kyiv-4, 01601, Ukraine\\
\vspace{1em}
E-mail: ivanova@imath.kiev.ua
\end{center}

\begin{abstract}
We investigate conservation laws of diffusion-convection equations to construct
first-order potential systems corresponding to these equations. We do two iterations of
the construction procedure, looking, in the second step, for the first-order conservation laws
of the potential systems obtained in the first step.
\end{abstract}

In her famous paper~\cite{Ivanova:Noether1918} Emmy Noether proved that
each Noether symmetry associated with a Lagrangian generates a conservation law.
(Modern treatment of  relationship
between components of Noether conserved vector and Lie--B\"acklund operators which are Noether
symmetries was adduced in~\cite{Ivanova:Ibragimov&Kara&Mahomed1998,Ivanova:Strampp1982}.
Kara {\it et al}~\cite{Ivanova:Kara&Mahomed2000,Ivanova:Kara&Qu2000} constructed conservation laws of some classes
of PDEs with two independent variables using
nonlocal symmetries.)
V.A.~Dorodnitsyn and S.R.~Svirshchevskii~\cite{Ivanova:Dorodnitsyn&Svirshchevskii1983}
(see also~\cite{Ivanova:Ibragimov1994V1}, Chapter~10)
completely studied the conservation laws for reaction-diffusion equations
\[
u_t=(d(u)u_x)_x+q(u).
\]
In~\cite{Ivanova:Anco&Bluman1996} a detailed discussion of the local nature of conservation laws derived from Noether's
theorem is presented.

A {\em conservation law} of a system of PDEs $\Delta(x,u_{(r)})=0$ is a divergence expression
${\rm div}\,F=0$ which vanishes for all solutions of these system.
Here $x=(x_1,\ldots,x_n),$ $u=(u^1,\ldots,u^m).$
$F=(F^1,\ldots,F^n),$ where $F^i=F^i(x,u_{(r)}),$ is a conserved vector of this conservation law
$u_{(r)}$ is the set of all the partial derivatives of the function $u$ with respect to $x$
of order no greater than~$r,$ including $u$ as the derivative of zero order.
We call the maximal order of derivatives explicitly appear in $F$ as the {\em order} of the
conservation law ${\rm div}\,F=0$.

The conservation laws are closely connected with non-local (potential) symmetries.
A system of PDEs may admit symmetries of such sort
when some of the equations
can be written in a~conserved form.
After introducing potentials for PDEs written in the conserved form as
additional dependent variables, we obtain a new (potential) system of PDEs.
Any local invariance transformation of the obtained system induces
a symmetry of the initial system.
If transformations of some of the ``non-potential'' variables
explicitly depend on potentials, this symmetry
is a non-local (potential)
symmetry of the initial system.
More details about potential symmetries and their applications
can be found in~\cite{Ivanova:Bluman88,Ivanova:Bluman89,Ivanova:bluman93}.

In this paper we consider conservation laws and potential systems for
nonlinear diffusion-convection equations of the form
\begin{equation} \label{eqf1}
u_t=(d(u)u_x)_x+k(u)u_x,
\end{equation}
that have a number of applications in physics
(see for instance~\cite{Ivanova:Barenblatt1,Ivanova:Barenblatt2,Ivanova:Kamin82}).
Here $d=d(u)$ and $k=k(u)$ are arbitrary smooth functions of $u,$ $d(u)\!\neq\! 0.$

The complete and strong group classification of~(\ref{eqf1}) was presented
in~\cite{Ivanova:Popovych&Ivanova2003NVCDCEsLanl} (see also references therein for more
details about symmetry analysis of classes intersecting class~(\ref{eqf1})).
The conservation laws associated with first-order potential systems for equations from class~(\ref{eqf1})
with $k=0$ were considered in~\cite{Ivanova:Doran-Wu1996}.
However, some of conservation laws and potential systems were missing there since the authors
did not investigate some necessary differential consequences of considered systems.

\begin{note}\label{noteeqtr}
If a local transformation connects two PDEs then under the action of this transformation a conservation law of
the first of these equations is transformed into a conservation law of the second equation, i.e. the equivalence
transformation establishes a one-to-one correspondence between conservation laws of these equations.
So, we can consider a problem of investigation of conservation laws with respect to the equivalence group
of the initial class.
\end{note}

Taking into account Note~\ref{noteeqtr}
we start our investigation of potential systems and conservation laws for the class~(\ref{eqf1})
from finding equivalence transformations.
Using the classical Lie approach, we find the Lie
invariance algebra of equivalence group $G^{\mathop{\rm \, equiv}}$
for class~(\ref{eqf1}). 

\begin{theorem}[see \cite{Ivanova:Popovych&Ivanova2003NVCDCEsLanl,Ivanova:Popovych&Ivanova2003PETsLanl}]
The Lie algebra of the equivalence group $G^{\mathop{\rm \, equiv}}$ for the class~\eqref{eqf1}
is
\[
A^{\mathop{\rm \, equiv}}=\langle
\partial_t ,\; \partial_x,\; \partial_u,\;
t\partial_t-d\partial_d-k\partial_k,\; x\partial_x+2d\partial_d+k\partial_k,\; u\partial_u,\;
t\partial_x-\partial_k\rangle .
\]
\end{theorem}

For class~(\ref{eqf1}) there also exists a nontrivial group of
discrete equivalence transformations generated by three involutive
transformations of the coupled alternation of signs in the sets of variables
$\{t,d,k\},$ $\{x,k\}$ and $\{u\}.$
It can be proved by the direct method that $G^{\mathop{\rm \, equiv}}$ coincides with
the group generated by the both continuous and discrete above transformations.

\begin{theorem}[see~\cite{Ivanova:Popovych&Ivanova2003NVCDCEsLanl,Ivanova:Popovych&Ivanova2003PETsLanl}]
Any transformation from $G^{\mathop{\rm \, equiv}}$ has the
form~
\[
\tilde t=\varepsilon_4t+\varepsilon_1, \quad
\tilde x=\varepsilon_5x+\varepsilon_7 t+\varepsilon_2, \quad
\tilde u=\varepsilon_6u+\varepsilon_3, \quad
\tilde d=\varepsilon_4^{-1}\varepsilon_5^2d, \quad
\tilde k=\varepsilon_4^{-1}\varepsilon_5k-\varepsilon_7,
\]
where $\varepsilon_1,$ \dots, $\varepsilon_7$ are arbitrary constants,
$\varepsilon_4\varepsilon_5\varepsilon_6\ne0.$
\end{theorem}

First we search the first-order conservation laws for equations from class~(\ref{eqf1}) in the form
\begin{equation}\label{conslaw}
D_tF(t,x,u,u_t,u_x)+D_xG(t,x,u,u_t,u_x)=0,
\end{equation}
where $D_t$ and $D_x$ are the operators of the total differentiation with respect to
$t$ and $x$ correspondingly,
namely, $D_t=\partial_t+u_t\partial_u+\cdots$, $D_x=\partial_x+u_x\partial_u+\cdots$.
Although this is the simplest class of conservation laws, it allows us
to deduce first-order potential systems
\begin{equation}\label{potsys1}
v_x=F,\qquad v_t=-G.
\end{equation}
Such systems as a rule admit a lot of nontrivial symmetries and so they are of a great interest.

\begin{definition}
Two conservation laws $D_tF+D_xG=0$ and $D_tF'+D_xG'=0$
are {\em equivalent} if
\[
F'=F+D_xH ,\qquad G'=G-D_tH,
\]
where $H$ is an arbitrary function of the same variables as the functions $F$ and $G$.
\end{definition}

\begin{theorem}
A complete set of inequivalent (with respect to the above equivalence relation) conservation laws
of the form~\eqref{conslaw} and potential systems~\eqref{potsys1} for
equations from class~\eqref{eqf1} is exhausted by the simple numbered cases of Table~1.
\end{theorem}

\begin{proof}
Let us substitute expression for~$u_{xx}$ deduced from~(\ref{eqf1})
into equation~(\ref{conslaw}) and decompose obtained equation with respect to $u_{tt}$ and $u_{tx}$.
As a result we have
\[
F=F^0(t,x,u,u_x),\qquad G=-F^0_{u_x}u_t+G^0(t,x,u,u_x).
\]

\newcounter{tbnIvanova} \setcounter{tbnIvanova}{0}
\begin{center}\small
{\bf Table~1.} Conservation laws and potential systems. \\[1ex]
\footnotesize\renewcommand{\arraystretch}{1.2}
\begin{tabular}{|l|c|c|c|c|l|}
\hline\raisebox{0ex}[2.8ex][0ex]{\null}
N & $d$ &$k$ &  $F$& $G$ & \hfill {Potential system \hfill} \\
\hline\raisebox{0ex}[2.8ex][0ex]{\null}
\refstepcounter{tbnIvanova}\thetbnIvanova&$\forall$ & $\forall$ & $u$ & $-du_x-\int k$ & $v_x=u,\ v_t=du_x+\int k$
\raisebox{0ex}[0ex][1.2ex]{\null}\\[-2.3ex]\multicolumn{6}{|c|}%
{\hspace*{-1.8ex}\dotfill\hspace*{-2ex}}\\[-1.05ex]\raisebox{0ex}[2.8ex][0ex]{\null}\thetbnIvanova.1&
$\forall$ & $0$ & $v$ & $-\int d$ & $v_x=u,\ w_x=v,\ w_t=\int d$
\\
\thetbnIvanova.2&$\forall$ & $d$ & $e^xv$&$-e^x\int d$
      & $v_x=u,\ w_x=e^xv,\ w_t=e^x\int d$
\\
\thetbnIvanova.3&$\forall$ & $\int d+ud$ & $e^v$ & $-e^v\int d$ & $v_x=u,
         \ w_x=e^v,\ w_t=e^v\int d$
\\
\thetbnIvanova.4&$u^{-2}$ & $0$ & $\sigma$ & $\sigma_vu^{-1}$ & $v_x=u,
\ w_x=\sigma,\ w_t=-\sigma_vu^{-1}$
\\
\thetbnIvanova.5&$u^{-2}$ & $u^{-2}$ & $\sigma e^x$ & $\sigma_vu^{-1}e^x$ & $v_x=u,
    \ w_x=\sigma e^x,\ w_t=-\sigma_vu^{-1}e^x$
\\
\thetbnIvanova.6&$1$ & $2u$ & $\alpha e^v$ & $\alpha_xe^v-\alpha u e^v$ & $v_x=u,
    \ w_x=\alpha e^v,\ w_t=\alpha u e^v-\alpha_xe^v$
\\[0.3ex] \hline\raisebox{0ex}[2.8ex][0ex]{\null}
\refstepcounter{tbnIvanova}\thetbnIvanova&$\forall$ & $0$ & $xu$ & $\int d-xdu_x$ & $v_x=xu,\ v_t=xdu_x-\int d$
\raisebox{0ex}[0ex][1.2ex]{\null}\\[-2.3ex]\multicolumn{6}{|c|}%
{\hspace*{-1.8ex}\dotfill\hspace*{-2ex}}\\[-1.05ex]\raisebox{0ex}[2.8ex][0ex]{\null}\thetbnIvanova.1&
$\forall$ & $0$ & $x^{-2}v$ & $-x^{-1}\int d$ &
     $v_x=xu,\ w_x=x^{-2}v,\ w_t=x^{-1}\int d$
\\[0.3ex]
\hline\raisebox{0ex}[2.8ex][0ex]{\null}
\refstepcounter{tbnIvanova}\thetbnIvanova&$\forall$ & $d$ & $(e^x+\varepsilon)u$ & $-(e^x+\varepsilon)du_x-
\varepsilon\int d$ &
     $v_x=(e^x+\varepsilon)u,\ v_t=(e^x+\varepsilon)du_x+\varepsilon\int d\!\!$
\raisebox{0ex}[0ex][1.2ex]{\null}\\[-2.3ex]\multicolumn{6}{|c|}{\hspace*{-1.8ex}\dotfill\hspace*{-2ex}}\\[-1.05ex]
\raisebox{0ex}[6.8ex][0ex]{\null}\thetbnIvanova.1 
&$\forall$ & $d$ & $\dfrac{e^x}{(e^x+\varepsilon)^2}v$
       & $-\dfrac{e^x}{e^x+\varepsilon}\int\!d$ &
      $\renewcommand{\arraycolsep}{0ex}\begin{array}{l}v_x=(e^x+\varepsilon)u,\ w_x=\dfrac{e^x}{(e^x+\varepsilon)^2}v,\\
       w_t=\dfrac{e^x}{e^x+\varepsilon}\int\!d  \end{array} $
\\[4.3ex] \hline\raisebox{0ex}[2.8ex][0ex]{\null}
\refstepcounter{tbnIvanova}\thetbnIvanova&$1$ & $0$ & $\alpha u$ &
$\alpha_xu-\alpha u_x$ & $v_x=\alpha u,\ v_t=\alpha u_x-\alpha_xu$
\raisebox{0ex}[0ex][1.2ex]{\null}\\[-2.3ex]\multicolumn{6}{|c|}%
{\hspace*{-1.8ex}\dotfill\hspace*{-2ex}}\\[-1.05ex]\raisebox{0ex}[5.6ex][0ex]{\null}\thetbnIvanova.1
& $1$ & $0$ & $\left(\dfrac{\beta}{\alpha}\right)_x v$
  & $-\alpha\left(\dfrac{\beta}{\alpha}\right)_x\! u-\left(\dfrac{\beta}{\alpha}\right)_tv$
   & $\renewcommand{\arraycolsep}{0ex}\begin{array}{l}v_x=\alpha u,
      \\[1ex]w_x=\left(\dfrac{\beta}{\alpha}\right)_x\! v,\quad
      w_t=\alpha\left(\dfrac{\beta}{\alpha}\right)_x\! u+\left(\dfrac{\beta}{\alpha}\right)_t\! v\end{array}$
\\[3.6ex]
\hline
\end{tabular}\end{center}\vspace{-1.mm}
{\footnotesize
Here $\varepsilon\in\{0,\pm 1\},$ $\int d=\int d(u)du$, $\int k=\int k(u)du$;
$\alpha(t,x)$, $\beta(t,x)$ and $\sigma(t,v)$ are arbitrary solutions of the
linear heat equation ($\alpha_t+\alpha_{xx}=0$, $\beta_t+\beta_{xx}=0$,
$\sigma_t+\sigma_{vv}=0$).
}
\vspace{2mm}

Taking into account the latter equations and splitting the rest of equation~(\ref{conslaw}) step by step with
respect to the powers of $u_t$ and $u_x$, we obtain a system of PDEs on functions
$F^0$ and $G^0$. Integrating these equations results in
\begin{gather}
F=D_x\varphi-\psi u,\qquad G=\chi-D_t\varphi+d\psi u_x
+\psi {\textstyle\int} k(u)du-\psi_x{\textstyle\int} d(u)du,\nonumber\\
\psi_t+d\psi_{xx}-k\psi_x=0,\label{eqconslaw1}
\end{gather}
where $\varphi=\varphi(t,x,u),$ $\chi=\chi(t),$ $\psi=\psi(t,x).$
With respect to the equivalence relation of conservation laws we can assume that $\varphi=\chi=0$.
Solving classifying equation~(\ref{eqconslaw1}) for different values of $d$ and $k$ we obtain Cases~1, 2, 3
and 4 of Table~1.
\end{proof}

\begin{note}
We can assume $(d,k)\not={\rm const}$ in Cases~1, 2 and~3.
(If $(d,k)={\rm const}$ Cases~1, 2 and~3 are included in Case~4 for different values of~$\alpha$.)
\end{note}


Let us consider first-order conservation laws of potential systems~1--4 from Table~1 which have the form
\begin{equation}\label{conslaw2}
D_tF(t,x,u,v,u_t,u_x,v_t,v_x)+D_xG(t,x,u,v,u_t,u_x,v_t,v_x)=0.
\end{equation}
These laws indeed nonlocal conservation laws for equations from class~(\ref{eqf1}).

\begin{lemma}\label{redconslaws}
Any conservation law of form~\eqref{conslaw2}
for any from systems 1--4 from Table~1 is equivalent to the law with the conserved density $F$ and the flux $G$
that are independent of the derivatives $u_t$, $u_x$, $v_t$ and $v_x$.
\end{lemma}

\begin{theorem}\label{theorconslaw2}
A complete set of inequivalent (with respect to the above equivalence relation) conservation laws
of the form~\eqref{conslaw2} for
equations from class~\eqref{eqf1} is presented in Table~1 with a~double numeration of cases.
\end{theorem}

\begin{note}
To prove Theorem~\ref{theorconslaw2} we use all functionally independent differential consequences
of correspondent potential systems. In Table~1 for the double numerated potential systems we omit equations
containing $v_t$ since they are only differential consequences of equations of these systems.
\end{note}

Note that potential symmetries arising for equations~(\ref{eqf1}) from Case~1 and~1.4 of Table~1 were found
by C.~Sophocleous~\cite{Ivanova:Sophocleous1996}.
The complete group classification of potential system corresponding to Case~1 of Table~1
was carried out in~\cite{Ivanova:Popovych&Ivanova2003PETsLanl}.

We plan to continue investigations of the subject. For the class under consideration we plan to
perform complete classification of potential and nonclassical (conditional) symmetries.
We also plan to study conservation laws of the more general structure (the high-order conservation laws
and conservation laws with pseudopotentials).

\vspace{-1ex}

\subsection*{Acknowledgements}

The author is grateful to
Profs.\ V.~Boyko, A.~Nikitin, R.~Popovych and  I.~Yehorchenko
for useful discussions and interesting comments.
This research was partially supported by National Academy of Science of Ukraine
in the form of the grant for young scientists.

\end{document}